# Non-equlibrium structure affects ferroelectric behavior of confined polymers

Daniel Martinez-Tong,[1] Alejandro Sanz,[2] Jaime Martín,[3] TiberioA Ezquerra[4] and Aurora Nogales[4]

1 Université Libre de Bruxelles, Boulevard du Triomphe, Bruxelles 1050, Belgium

2 DNRF Centre "Glass and Time", IMFUFA, Department of Sciences, Roskilde University, Postbox 260, DK-4000 Roskilde, Denmark

3 Department of Materials and Centre of Plastic Electronics, Imperial College London, London SW7 2AZ, UK

4 Instituto de Estructura de la Materia IEM-CSIC, C/ Serrano 121, Madrid 28006, Spain

**Abstract**

The effect of interfaces and confinement in polymer ferroelectric structured is discussed. Results on confinement under different geometries are presented and the comparison of all of them allows to evidence that the presence of an interface in particular cases stabilizes a ferroelectric phase that is not spontaneously formed under normal bulk processing conditions

## Introduction

Polymers are hierarchical systems, with structure in different length scales. Most macroscopic properties of polymers have their origin in physical phenomena associated with length scales of the order of nanometers. Hence, confining polymers to the nanoscale may end up in a material with different physical properties. This fact is remarkably important nowadays with the development of polymer applications in devices with smaller and smaller sizes. From the fundamental viewpoint, for several decades experiments have been designed to evaluate the role of limitation of space in several physical processes, like the glass transition of polymers(Alcoutlabi and McKenna, 2005), changes in the dynamics, either local or segmental (Priestley et al., 2015), among others. However, in those designed experiments, besides the pure finite size effects, another issue has to be inevitably considered, and that is the presence of interphases, that become extremely important in confinement experiments, when the characteristic size is in the length scale of the interaction of the system with a substrate. This aspect has drawn the attention of many researchers recently, and the presence of adsorption layers, and how this layer can be modified and tuned to lead to different polymer properties is an area of intense research nowadays (Xu et al., 2014, Jiang et al., 2014, Napolitano et al., 2012, Napolitano and Wubbenhorst, 2011)

The presence of interfaces, either with solid substrates or free surfaces, modify parameters such as number of entanglements, chain orientation, chain packing, local chain mobility and glass transition temperature ($T_g$), mechanical properties, crystallization kinetics or crystals orientation (Vanroy et al., 2013, Napolitano and Wubbenhorst, 2006, Capitan et al., 2004, Martin et al., 2014, Martín et al., 2013, Sun et al., 2013), from their bulk counterpart.

Semicrystalline polymeric materials exhibit a rich variety of morphologies (Bassett, 1981) [1] which are responsible, to a high extent, for the macroscopic properties of the manufactured polymer product. For example, depending on the morphology, chemically equivalent polymers can generate materials with very different macroscopical moduli (Wilkinson and Ryan, 1999). [2] Therefore, the resultant morphology is as discriminating as to the attainment of specific properties as the chemical configuration. The presence of foreign interfaces is one of those aspects that modify extensively the morphology in semicrystalline polymers. In some particular cases, when the polymer exhibits more than one crystalline form, the presence of interfaces can favor the development of one of high energy crystalline forms that are not the most stable in the absence of external interfaces (Balik and Hopfinger, 1980, Hua et al., 2007). This effect has been used by the industry to design additives that nucleate a particular crystalline form, like the β nucleants in isotactic polypropylene (Mollova et al., 2013, Gahleitner et al., 2012, Gahleitner et al., 2011). In other cases, confined at the nanoscale level, polymers crystallize much slower than in bulk, and sometimes, the formation of ordered structures results inhibited for extremely long experimental time scales.(Capitán et al., 2004, Vanroy et al., 2013, Martínez-Tong et al., 2014)

One of the bulk properties that is directly related to the polymer crystalline morphology and crystal form is ferroelectricity in Poly(vinylidene fluoride) (PVDF) and its copolymers with Trifluoroethylene (P(VDF-TrFE). In general, ferroelectrics are polar substances of either solid (inorganic or polymeric) or liquid crystal, in which spontaneously generated electric polarization can be reversed by inverting the external electric field (Horiuchi and Tokura, 2008). The critical electric field for reversing the polarization is called the coercive field. The electric displacement (D) as a function of field strength (E) consequently draws a hysteretic curve (D–E loop) between opposite polarities, and this electric bistability can be used, for example, for non-volatile memory elements (Youn Jung et al., 2010, Zhu and Wang, 2012, Yang et al., 2013). The ferroelectric compounds usually have a Curie temperature $T_C$ for a ferroelectric-to-paraelectric phase transition. As the temperature approaches $T_C$, the dielectric constant (κ), obeying the Curie–Weiss law, is amplified to large values, which can be exploited for a high-κ condenser. Specifically, P(VDF-TrFE) copolymers with TrFE content higher than ≈ 10 % crystallize in a ferroelectric phase (Yagi et al., 1980b, Zhu and Wang, 2012), monoclinic or orthorhombic, which is similar to the β-crystalline and ferroelectric phase of PVDF (Lovinger, 1983). This ferroelectric phase is stable at room temperature and suffers a phase transition to a paraelectric one, hexagonal, as temperature increases (Baltá-Calleja et al., 1993, Yang et al., 2013). This transition temperatur called Curie temperature, increases with TrFE content (Baltá-Calleja et al., 1993).

Confinement and interfaces are particularly important issues for these systems, since arrays of polymeric nanostructures showing ferroelectric and piezoelectric behavior, have a well-recognized potential for the fabrication of miniaturized and novel organic electronic devices, such as high density nonvolatile memories, (Chuanqi et al., 2009, Martinez-Tong et al., 2013) low voltage operation polymer capacitors (Jung et al., 2009b), nanopressure sensors (Mandal et al., 2011), among other applications (Cauda et al., 2015). However, the use of ferroelectric materials in microelectronics is, however still not very wide because the continuous downscaling of electronics requires to overcome the technological drawback because of the difficulty on stabilizing ferroelectricity on the nanoscale.

It has been reported how, by imposing confinement and interaction with the walls of an alumina template, PVDF homopolymer is able to crystallize in the polar γ crystalline form, when the system is processed by solution wetting. In this case, the interaction between polymer chains and the porous membrane's walls imposes a flat-on lamella growth along the nanorods long axis, while improving crystal orientation (García-Gutiérrez et al., 2010).

The presence of interfaces on the dynamics of PVDF confined in alumina nanopores has been also investigated by dielectric spectroscopy (Martín et al., 2009). A strong deviation of the relaxation behavior of PVDF embedded within the nanopores is observed as compared to that of the bulk. When the restriction in size is larger, that is, when the pore diameter is comparable with the size of the adsorbed layer, the existence of a highly constrained relaxation associated with the polymer-alumina interfacial layer is observed.

In the above mentioned works, the presence of interfaces induces the crystallization of PVDF in a polar phase. P(VDF-TrFE) however, spontaneously crystallizes in the β polar phase, and they are ferroelectric at room temperature under normal processing conditions. The question is whether confinement and interfaces modify this intrinsically ferroelectric phase.

To shed some light into that question, we discussed here the behavior of ferroelectric copolymers P(VDF-TrFE) under different confinement conditions, where the dimensionality of the confinement has been modified, from 1D confined (thin films), 2D confined (nanorods) or 3D confined (nanospheres).

## Short review on the crystalline structure, thermal properties and morphology of bulk ferroelectric polymers

Copolymers of PVDF with trifluoroethylene are semicrystalline systems in which the crystalline part exhibits ferroelectricity. For TrFE molar fractions below 85%, this crystalline phase suffers a phase transition below the melting point. TrFE units are included in the unit cell of PVDF and enlarge it since they are bulkier. This enlargement weakens the intermolecular interactions among chains, and destabilizes the ferroelectric phase with the all-trans conformation. The more TrFE segments in the chain, the more unstable the ferroelectric phase becomes, and as a result the ferroelectric-to-paraelectric Curie transition temperature ($T_C$) decreases with the increase of TrFE units (Zhu and Wang, 2012). This change in the crystalline structure, from the pseudohexagonal phase where parallel chains are in all-trans planar conformations to the hexagonal paraelectric phase can be easily observed by differential scanning calorimetry (DSC) where an endothermic peak before the melting temperature is observed. The temperature of this transition is referred to as the Curie Temperature ($T_C$). Figure **1**(a) shows the DSC traces of PVDF-TrFE with different molar content. Two main endothermic peaks appear. The one at lower temperature corresponds to the ferro-para phase transition (F-P transition). The position of this peak is the Curie temperature. As observed, $T_C$ is higher for the smaller TrFE content. The endothermic peak appearing at higher temperature is the melting of the paraelectric phase.

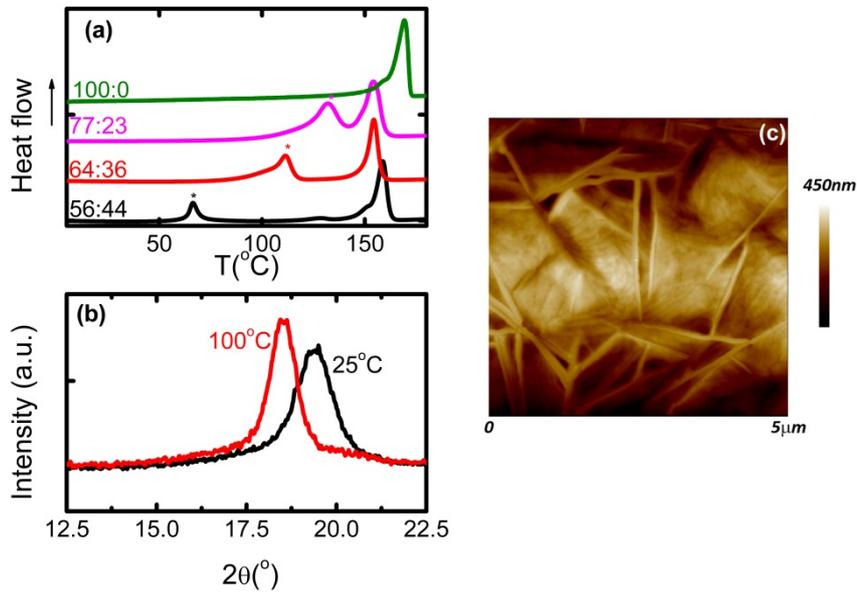

**Fig. 1. (a)** Differential Scanning Calorimetry traces obtained at 20º/min, for different TrFE molar contents, as indicated in the label. The star indicates the Curie temperature. For comparison, the DSC trace showing the melting of the non ferroelectric α phase of the PVDF homopolymer (labelled as 100:0) has been included**. (b)** WAXS measurements at 25 ºC (black) and 100ºC (red) for P(VDF-TrFE) 56:44 copolymer. Both temperatures are below and above the Curie temperature respectively. (c) AFM topography image of a P(VDF-TrFE) 56:44 film of thickness ≈ 200 nm

At room temperature, Wide Angle X Ray Scattering (WAXS) patterns of bulk P(VDF-TrFE) show a sharp maximum in the region around $2\theta \approx 20º$, characteristic of the *200,110* reflections of the orthorhombic ferroelectric crystalline phase of P(VDF-TrFE). The exact position of these reflections depend slightly on the molar content of TrFE (Lovinger, 1983). As an example, Figure **1**(b) shows the WAXS pattern corresponding to the 56:44 PVDF-TrFE copolymer at room temperature. However, at temperatures above $T_C$, WAXS pattern exhibit also an intense and sharp maximum, but at a different position, particularly at $2\theta \approx 18.5º$ for the 56:44 P(VDF-TrFE) copolymer.

From the morphological viewpoint, P(VDF-TrFE) copolymers are characterized by random large needlelike crystals (Jung et al., 2009a). This morphology has been associated to edge-on lamellae, that also coexist with flat-on crystals appearing as irregular flakes on the AFM image, Figure 1(c).

In the vicinity of $T_C$, the dielectric constant shows a strong increase. This increase can be observed by broadband dielectric spectroscopy, which also provides an excellent method for investigating the molecular dynamics in ferroelectric copolymers below and above the $T_C$. Figures 2(a) and (b) show the dielectric permittivity and dielectric loss values respectively as a function of temperature for particular frequencies. At low temperatures a maximum in $\varepsilon''(T)$ around -40ºC is observed. This can be attributed to the segmental relaxation associated to the glass transition in agreement with previous reports. (Yagi et al., 1980b, Ngoma et al., 1991, Ezquerra et al., 1994). In this type of polymers the segmental relaxation has been labeled as β-relaxation since dynamical mechanical analysis shows another relaxation, labeled originally as α, associated to the crystalline phase appearing at higher temperatures (Yagi et al., 1980b, Ngoma et al., 1991). In order to comply with the literature we adopt here a similar criterion.

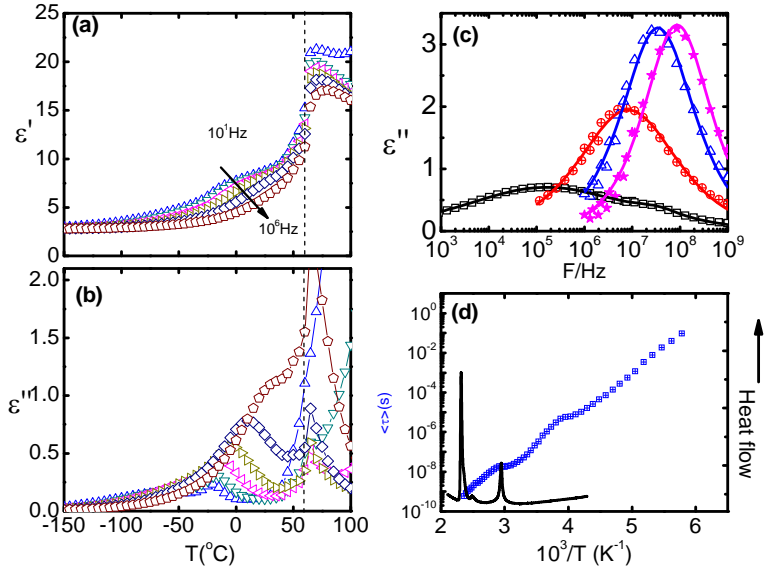

**Fig. 2.** Dielectric spectroscopy of bulk 56:44 P(VDF-TrFE) (a) dielectric constant ε' at various frequencies. (b), dielectric losses ε'' at various frequencies. discontinuous line indicates the $T_C$. (c) ε'' as a function of frequency for various temperatures: 0ºC (□), 50 ºC (⊗), 100 ºC (△) and 125 ºC (★). (d) Relaxation map for the 56:44 PVDF-TrFE copolymer, with the average relaxation time obtained from the position of the maximum loss as $<\tau>=(2\pi f)^{-1}$ (□) and the DSC trace (continuous) line.

Figure 2.c shows isothermal dielectric loss, ε'', data as a function frequency for bulk PVDF-TrFE 56:44 copolymer. In isothermal plots, the β relaxation appears as a maximum that at T=0ºC is centered on $10^5$Hz, and whose position shifts toward higher frequencies as temperature is increased. A dramatic increase of the intensity of the relaxation is observed as the F-P transition is approached, above 60 ºC for the 56:44 copolymer. It has been proposed that for these type of copolymers the intense relaxation observed centered around $10^7$-$10^8$ Hz appears as a consequence of the onset of rotational motion of the molecules within the paraelectric crystalline phase (Ezquerra et al., 1994, Yagi et al., 1980b). This process coexists with the segmental relaxation, β, occurring in the amorphous phase. The relaxation map of this system is presented in Figure **2**.d, together with the DSC trace for the 56:44 P(VDF-TrFE) copolymer. As observed, at low temperature there is a clear Arrhenius tendency. In this region, local motions are responsible for the maximum observed in the dielectric loss as a function of temperature. Following it, a Vogel Fulcher Tamman behavior of the relaxation time is observed, in the region where the segmental relaxation becomes relevant. And, around $10^3/T=3$ K$^{-1}$, (T ~ 60ºC) a clear kink is observed in $<\tau>$ that corresponds to the crystalline transition from the ferroelectric to the paraelectric phase. At higher temperatures, the most important contribution to the dielectric losses is the relaxation associated to the rotational dynamics in the paraelectric crystalline phase, and as it is observed in figure 2.d, it follows an Arrhenius behavior. This trend is reproduced for other copolymer molar fractions, with the kink in $<\tau>$ occurring at each $T_C$.(Yagi et al., 1980a).

## Non equilibrium effects in ferroelectric polymers confined in 1, 2 and 3 dimensions.

Confinement experiments are complex ones, particularly in polymers, since many aspects have to be considered. Besides the existence or absence of interfaces, the geometry itself plays an important role. 1D confinement geometry (thin films) and 2D confinement (nanotubes and nanorods) implies that the polymer chain has to accommodate in an anisotropic fashion, whereas in 3D (nanospheres) the confinement is isotropic. In each of the above mentioned experiments, interfaces have to be considered. A particular case of 1D confinement, which is the case for supported thin films, it has been demonstrated that there is a strong impact of interfaces on the static and dynamical properties of the polymer (Napolitano et al., 2013). These interfaces also appear in the case of polymer confined in cylindrical pores. Tanaka and coworkers demonstrated that, the slower dynamics near a wall is induced by wall-induced enhancement of 'glassy structural order', which is a manifestation of strong interparticle correlations (Watanabe et al., 2011). The presence of the solid interface favours the presence of clusters with a preferential bond orientational order. When the polymer system under these circumstances is semicrystalline, the crystallization process is modified in two different ways. On one hand, the slower dynamics due to the polymer-chain interactions delays the crystallization process (Vanroy et al., 2013, Martínez-Tong et al., 2014). On the other hand, the development of clusters with preferential orientation templates the crystallization in particular directions, (Martin et al., 2014, Martín et al., 2013, Garcia-Gutierrez et al., 2013). In the following subsections, non-equilibrium effects in the ferroelectric character of P(VDF-TrFE) copolymers, due to confinement and to the presence of interfaces will be discussed. The case of 1D confinement on supported thin films, 2D confinement on alumina cilindrical nanopores and 3D confinement in free standing nanospheres will be considered.

### *1D confinement. PVDF-TrFE Thin films.*

The interaction with an external surface in any material may induce a discontinuity and anisotropy in the energetic balance of the system, imposing effects different from those observed in the bulk material. In ferroelectric materials, it has been observed that finite size effects may appear. However, it was probed that ultrathin films of P(VDF-TrFE) copolymers prepared by the Langmuir Blodgett (LB) technique exhibit a F-P transition with a $T_C$ nearly equal to the bulk value, but, in addition, another transition appears at a lower temperature, associated with the surface layers only (Bune et al., 1998). LB films are highly crystalline, therefore, interfacial layers in this particular case do not present a large number of deffects. Nevertheless, dielectric experiment have reported the presence of a dielectric segmental relaxation in P(VDF-TrFE ) LB films, attributed to the motion of amorphous chains.(Asadi et al., 2008)

Polar interfacial phases are also reported in PVDF homopolymer LB films (Wang et al., 2012). It was found that by LB deposition direct formation of ferroelectric β phase in PVDF with the molecular chains parallel to the substrates and the dipoles aligned perpendicular to the substrates. These experiments are clear evidence on how the presence of an interface favors a different crystallographic phase, which in bulk represents a higher energy one, and that is stabilize thanks to the interaction with the surface.

The dielectric properties of ferroelectric P(VDF-TrFE) thin films prepared by spin coating have not been widely studied. It has been reported (Martin et al., 2012a) that the dielectric behavior of P(VDF-TrFE) spin coated thin films is similar to that of the bulk system. P(VDF-TrFE) polymers prepared in this way crystallize in the polar β form, as the bulk system does (Park et al., 2008b, Park et al., 2008a). The preferential orientation of the crystal depends highly on the interaction with the substrate (Lee et al., 2010, Park et al., 2008b, Park et al., 2008a) , and because of that, in thin films prepared in this way, the control of the polarization is difficult due to the strong heterogeneity of the crystalline morphology leading to non-uniform local fields and non-uniform spreading of the ferroelectric domains. To overcome these spreading, several methods have been reported to stop the spreading of the ferroelectric domains, by confining the crystallization to limited space.(Nougaret et al., 2014, Garcia-Gutierrez et al., 2013, Piraux et al., 2011, Hu and Jonas, Martinez-Tong et al., 2013).

## *PVDF and ferroelectric PVDF-TrFE confined into tubes and rods (2D confinement)*

Several types of nanostructures have been employed to study the confined crystallization of polymers. The system composed of polymers confined into anodic aluminum oxide (AAO) nanopore arrays has been widely used because of its high tunability on the degree of confinement in terms of the pore diameter, a mechanical rigidity of the hard pore walls, and a well-defined confining geometry.(Martin et al., 2012b). High surface energy solids such as metal oxides are wettable by almost all low surface energy systems like polymer melts. When a polymer melt is deposited on top of a porous anodic aluminum oxide (AAO) template, the polymer wets the pores. Depending on the size of the pore, the annealing temperature, and the spreading coefficient for the given polymer/alumina system, two wetting regimes (partial and total wetting) are observed, giving rise to two different polymeric nanostructures inside the pores (nanorods or nanotubes) (Steinhart, 2002).

For the preparation of polymer nanorods or nanotubes confined into alumina templates, several polymer pellets were placed onto the AAO templates in order to carry out the infiltration by ~~the~~ melt wetting. Figure **3**.a, shows Atomic Force Microscopy (AFM) images from one of the AAO templates. The molten polymer wets the surface of the AAO template by forming a precursor film. Depending on wether the radius of the pore is larger or smaller than the thickness of that precursor film, either a complete filling is obtained (polymer nanorods) or a partial filling (polymer nanotubes). The AAO with the polymer pellets were then annealed at 240ºC under a nitrogen atmosphere for 45 min. After the infiltration process, the samples were quenched under ice water. In order to force the separated crystallization of each individual nanorod, the residual polymer film located on the AAO surface, which connected the nanostructures, was removed with a blade. Finally, the nanorods, now not connected by the residual film, were again molten at 240ºC and quenched in ice-water.

Figure **3**.b. shows Scanning Electron Microscopy pictures obtained at two different magnifications of the nanostructures of PVDF, removed from the template.

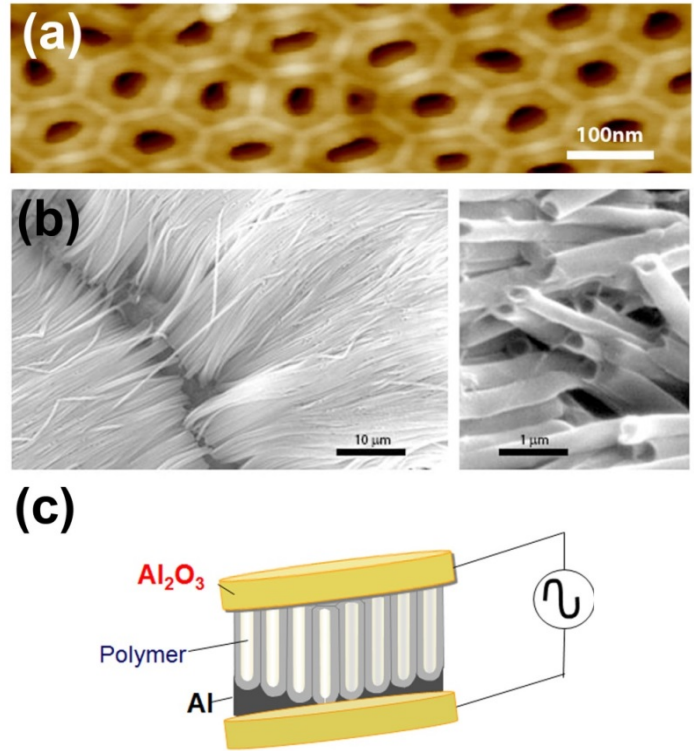

**Fig. 3.(a)** AFM image of an AAO template surface. (b) SEM image of PVDF nanotubes at different magnifications after being removed from the AAO template.(c) Schematic drawing of the capacitor ensemble employed for dielectric characterization.

Dielectric measurements were performed at selected frequencies and temperatures. Figure **3**.c. present and sketch of the measuring setup used. With this measuring set-up, the dielectric signal recorded has a contribution from the polymer nanostructures and from the alumina. However, the contribution from alumina is independent of temperature, as it was proved by performing the dielectric experiment on an empty template. Therefore, the changes in the dielectric signal observed as a function of temperature can be attributed solely to the polymer nanostructures.

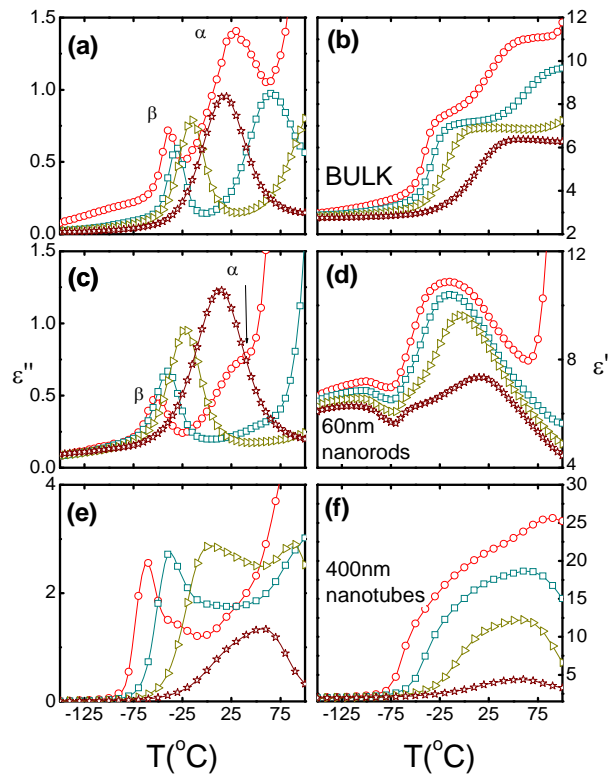

**Fig. 4.** Dielectric Loss (ε")(a,c and e) and Dielectric Permitivitty (b,d and f) (ε') for the different studied PVDF nanostructures and comparison with the bulk behavior at different frequencies. (O) $10^0$Hz, (□) $10^2$Hz, (▽) $10^4$Hz and (☆) $10^6$Hz.

Figure **4** presents the obtained values for the dielectric loss (left) and the dielectric permittivity (right) as a function of temperature for Bulk PVDF and PVDF nanostructures, precisely, nanorods of 60nm in diameter and nanotubes of 400nm in diameter. Dielectric loss behavior for bulk PVDF, exhibits two main relaxation processes in the studied frequency-temperature window. The one appearing at lower temperatures (β relaxation) can be attributed to the segmental motions of amorphous PVDF chains above the glass transition temperature. The maxima in dielectric loss at higher temperatures, that consistently with the literature we have labeled it as α, is attributed to local motions in the crystalline phase of PVDF. The β and α relaxation processes are also appreciable in the dielectric loss spectra for the 60nm rods. Previous results show that this behavior persists even in smaller nanorods down to 35nm in diameter (Martín et al., 2009). In that work has been shown that for smaller rod diameter, the behavior of ε" changes dramatically and it can be associated to the existence of a highly constrained relaxation associated with the polymer-alumina interfacial layer (Martín et al., 2009). This sort of highly constrained behavior is also present in the PVDF nanotubes of 400nm in diameter. Despite the higher diameter of these nanostructures, the results can be understood considering that the wall thickness of the nanotubes is also of the order of 20nm (see Figure **3**), and therefore, the entire dielectric signal is associated with interfacial PVDF. This interfacial PVDF must be also present in nanorods with higher diameter (60nm), although it is difficult to detect observing only the values of ε". However, dielectric permittivity (ε') values show strong changes compare to those of the bulk for all the nanostructures studied. Rods of 60 nm diameter presents a sudden increase of ε' at around -75ºC, which more or less corresponds to the increase observed in bulk PVDF and can be associated in a first attempt to the segmental β relaxation, but, instead of the typical step like behavior, in the 60nm nanostructure ε' exhibit a very broad maximum. This behavior is not similar to that of a F-P transition, since, unlike the dielectric peak associated to that situation, here the maximum in ε' shifts progressively toward higher temperature with frequency. This is a feature that has been observed in relaxor ferroelectrics in general (Yuan et al., 2005) and in PVDF based ferroelectric polymers chemically modified or irradiated (Bobnar et al., 2003). As mentioned before, recently, X ray micro diffraction studies have proved that, it is possible to obtain arrays of isolated polar nanorods of PVDF by solution wetting nanoporous alumina template, and that the polymer crystallized into the polar γ-phase. This peculiar crystallization is attributed to confinement effects and interaction of the polymer solution with the alumina walls during wetting (García-Gutiérrez et al., 2010). The dielectric results presented in figure 4 indicate that these polar structures are also present in PVDF nanorods prepared by melt wetting the alumina template, and therefore, is the interaction with the alumina wall the responsible for the stabilization of a polar phase in PVDF. This interpretation is also in line with recent reports where the (Zhu et al., 2014) arrangement of the first layers in the homopolymer PVDF in LB films give rise to crystallization in the polar phase.

Distortion of the bulk ferroelectric character is also observed in nanostructures of P(VDF-TrFE) copolymers. Figure 5 shows ε' and ε" values obtained for nanorods of P(VDF-TrFE) 56:44 molar fraction with two different diameters, 60nm and 35nm and the dielectric response of the same system in the bulk for comparison. There is a strong increase of ε' in the nanorods at temperatures around -50ºC. This temperature region corresponds to that of the segmental relaxation region of the bulk polymer, and therefore, can be attributed to the β relaxation. In order to observe further differences in the nanorods and the bulk dynamics, in Figure 6.a and b show ε' and ε'' values for P(VDF-TrFE) copolymer confined in 60nm and 35 nm diameter pores, together with the values corresponding to the bulk, normalized to the value at the maximum located around T=-40ºC. As observed, the β relaxation, associated to the segmental motion above $T_g$, is clearly visible in the confined systems. However it appears slightly shifted towards lower temperatures, which is compatible with a lower crystallinity due to confinement (Nogales et al., 2000). The main difference in the behavior of the bulk and nanopore confined system appears in the vicinity of the bulk $T_C$. As mentioned before, associated to the Curie transition there is a strong increase in ε' in the bulk ferroelectric copolymer. However, this increase is not observed in the confined polymer into alumina templates (Figure **5**). Instead, ε' exhibit a relaxation like behavior, where ε' decreases when increasing the temperature.

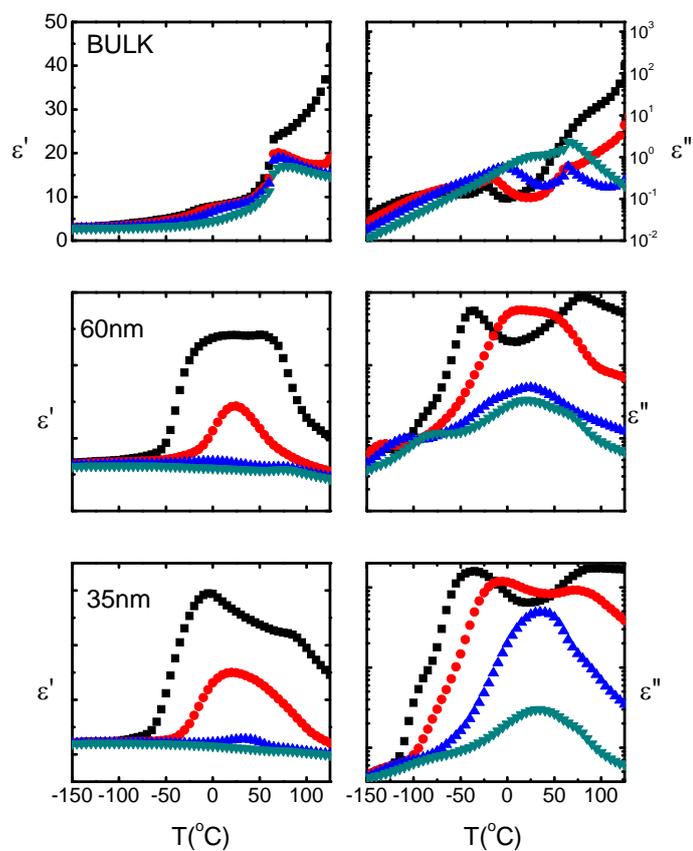

**Fig. 5. .**( Left) Dielectric Permitivitty (ε') and (Right) Dielectric Losss (ε'') values as a function of temperature for different diameter PVDF-TrFE nanorods, as indicated by the label and comparison with the bulk behavior at different frequencies. (■) $10^0$Hz, (●) $10^2$Hz, (▲) $10^4$Hz and (◆) $10^6$Hz. Values for the 60nm and 35 nanorods should be considered as relative, since the geometric factor cannot be calculated in this measuring geometry

At temperatures around $T_C$, the confined system shows different features, like a maximum in both ε' and ε" and the absence of the strong increase associated to the F-P transition. These results may indicate that there is an inhibition of the F-P transition in the ferroelectric polymer confined into alumina nanopores. DSC results also support this idea. In figure **6**.c the DSC traces for P(VDF-TrFE) confined into pores with 60nm and 35nm in diameter is presented, and compared with that of the bulk. As observed, in the 60nm confined sample, the endothermic F-P peak appears shifted towards higher temperatures, very weak and broad. In the 35nm confined system, the peak is not observed.

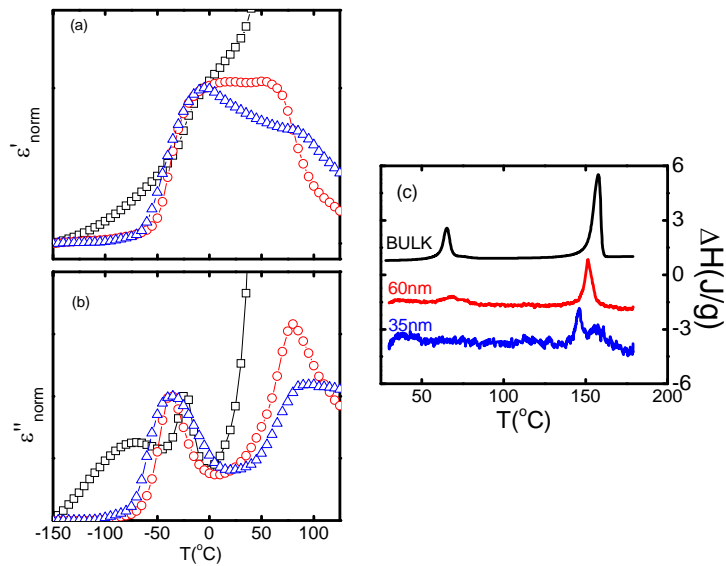

**Fig. 6**. (a) Dielectric Permittivity and (b) Dielectric Loss as a function of frequency for the bulk 56:44 P(VDF-TrFE) and different diameter nanorods confined in alumina templates (O)60nm rods and (△) 35nm rods at $F=10^4$Hz. Values are normalized to the value of ε' and ε" at the maximum of the β relaxation (T around -50ºC). (c) Differential scanning calorimetry of the bulk 56:44 PVDF-TrFE polymer and for the nanorods of 60 and 35 nm respectively.

Previous experiments performed on ferroelectric copolymers under confined geometries (Bune et al., 1998) have shown that ferroelectricity is preserved, but the F-P transition is broadened and, in particular cases, an interfacial F-P transition is observed. In particular, highly ordered Langmuir Blodgett multilayers of PVDF-TrFE (70:30) exhibit an interfacial ferroelectric transition located around 20ºC.(Palto et al., 1996)

The results described indicates that confining PVDF and P(VDF-TrFE) copolymers in cylindrical nanopores formed in alumina templates induces severe changes in the ferroelectric behavior of the polymer. In the case of the PVDF homopolymer, confinement into alumina template produces an enhancement of fer-roelectric like features in the dielectric spectroscopy experiments that might indicate the formation of an interfacial ferroelectric phase. In the case of the P(VDF-TrFE) copolymers, confinement into alumina templates produces an inhibition of the ferro-para transition. In order to check wether this inhibition is due to purely spatial confinement or to the interaction with the confining walls, confined P(VDF-TrFE) was prepared in the form of nanospheres, where no interaction with walls is present. The results are discussed in the following section.

## *Ferroelectric PVDF-TrFE confined in the form of nanospheres (3D confinement)*

**Preparation of the nanospheres**

P(VDF-TrFE) nanoparticles were prepared by the dialysis nanoprecipitation method (Zhang et al., 2012). The polymer was dissolved in *N,N*-dimethyl acetamide (DMA) at room temperature under continuous stirring. A dialysis membrane, with molecular weight cutoff 10–12 kg/mol (Visking), was carefully washed in water, soaked in DMA and left inside a glass container with the solvent for 30 min. This ensures the complete rinse of the membrane on the solvent eliminating water and avoiding macroscopic precipitation. Afterwards, the polymer solution was poured into the membrane and subsequently, the membrane was submerged and dialyzed against 2 L of distilled water under mild stirring. This allows the exchange of solvent/water molecules giving rise to the nanoprecipitation. During the first 2 h, the dialysis solution was replaced with fresh distilled water every 30 min. Subsequently, for the next 4 h, the water was replaced every hour. Dialysis nanoprecipitation was allowed to take place overnight, resulting in a total exchange time of about 20 h. The final result is a white aqueous emulsion of surfactant free nanoparticles inside the membrane, where no DMA solvent is left. The nanoparticles were deposited onto a silicon wafer by spin coating, in order to evaluate their size distribution by AFM (figure **7**)

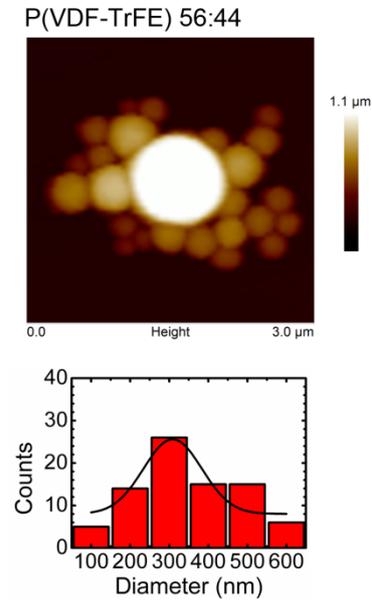

**Fig. 7.** AFM images and histograms for P(VDF-TrFE) nanoparticles. Histograms represent the diameter distributions of the nanoparticles.

The dielectric loss values measured as a function of frequency for the nanoparticles obtained in that way show features that are very similar to the ones obtained for the bulk material. In figure 8, ε' and ε'' data as a function frequency are presented.

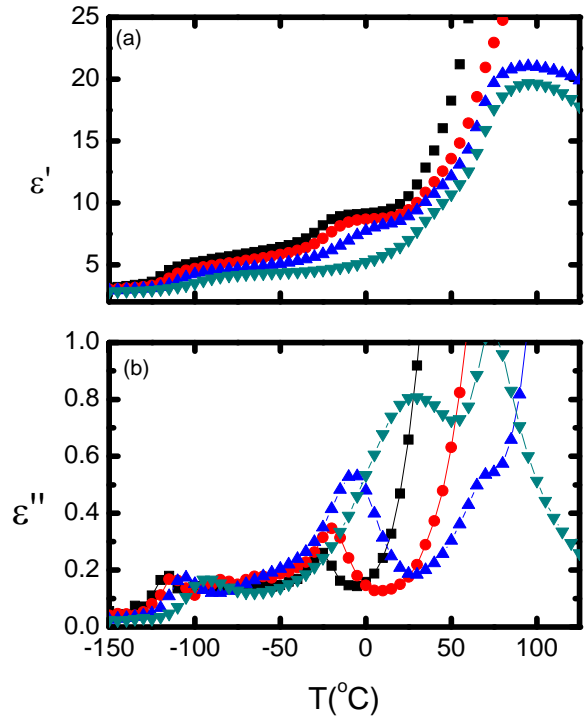

**Fig. 8.** (a) Dielectric permittivity and (b) dielectric loss as a function of temperature for various frequencies obtained from 56:44 P(VDF-TrFE) nanospheres. (■)$10^0$Hz, (●)$10^2$Hz, (▲)$10^4$Hz(▼)$10^6$Hz

Comparing figures **2** and **8**, the only difference appears at very low temperature, with the presence of a relaxation peak in ε'' located around -125ºC for $10^0$Hz The maximum of this relaxation shifts towards higher values as temperature increases and the relaxation becomes sharper. Similar with what it has been observed in the dielectric experiments for the bulk system, a dramatic increase of the intensity of the relaxation is observed as the ferroelectric-paraelectric transition is approached ($T_C$ approx. 50ºC for the 56:44 copolymer). This process coexists with the segmental relaxation, β, occurring in the amorphous phase.

The dielectric behavior in the ferroelectric polymer nanoparticles is therefore, very similar to that of the bulk and considerably different from the observed in the case of the ferroelectric polymer nanorods confined into alumina cylindrical pores, particularly in the small diameter case. For the nanoparticles, there is no indication of any interfacial relaxation.

In order to evidence these differences, figure 9 shows ε'' for the bulk, the 60nm and 35 in diameter rods in alumina template samples and the nanospheres measured at high frequency. As observed, the relaxation associated to the paraelectric phase is clearly visible in the nanospheres and also in the case of the 60 nanometer in diameter rods. However, the relaxation is absent in the case of the 35nm in diameter nanorods, which supports the interpretation of inhibition of the F-P transition in these highly constrained structures.

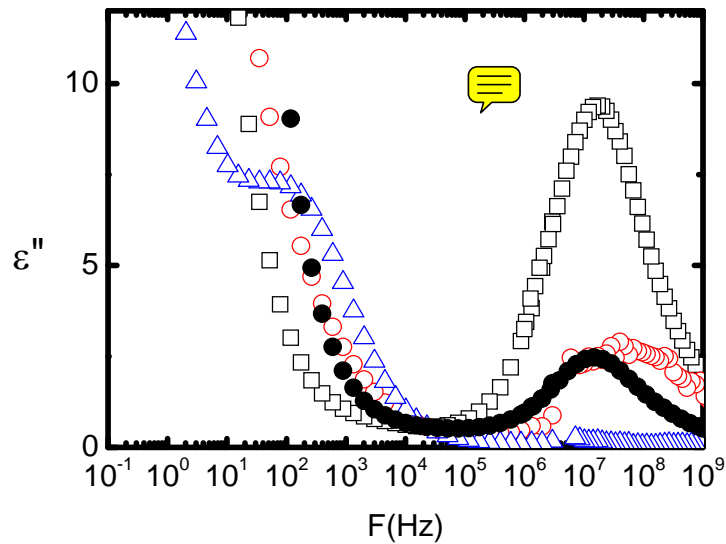

**Fig. 9.** ε'' as a function of frequency for the confined system studied in this work Values obtained for the bulk are also presented for comparison (□) Bulk, (O,△) Nanorods in Alumina pores of 60 and 35 nm diameter respectively. (●) Nanospheres.

## Conclusions

Ferroelectric properties in PVDF and P(VDF-TrFE) copolymers are directly associated with their crystalline phase. Confinement may induce certain modification to the crystalline phase. However, the observe changes in the ferroelectric behaviour cannot be solely attributed to spatial confinement. It is the presence of interface with whom the polymer chains interacts, the responsible for the appearance of peculiar ferroelectric behavior. To name some of the ones discussed in this chapter, when PVDF is confined into alumina cylindrical templates, an interfacial layer appears, that present ferroelectric like behavior, whereas the bulk of this homopolymer is paraelectric under normal processing conditions. However, in P(VDF-TrFE) copolymers confined in alumina templates. For severe confinement, the transition from ferroelectric to paraelectric is inhibited, indicating that the interaction with the confining wall stabilizes de ferroelectric phase. Confinement without interfaces, like in the case of polymer nanoparticles, do not change the ferroelectric behavior of the polymer. It may decrease the crystallinity of the system but the crystals, that the ferroelectric character of the polymer, have the same properties as in the bulk.